\documentclass[10pt,letterpaper]{article}

\usepackage{cogsci}
\usepackage{pslatex}
\usepackage{apacite}

\usepackage{epsfig}
\usepackage{graphicx}
\usepackage{times}
\usepackage{amsmath}
\usepackage{amssymb}
\usepackage{caption}
\usepackage{subcaption}
\usepackage{float}
\usepackage{ulem}
\usepackage{wrapfig}
\usepackage{color}
\usepackage{tikz}
\usepackage{gensymb}

\title{Characterizing the Temporal Dynamics of Information in Visually Guided Predictive Control Using LSTM Recurrent Neural Networks}

\author{{\large \bf Kamran Binaee (kb4000@rit.edu)}\AND {\large \bf Anna Starynska (as3279@rit.edu)}\AND {\large \bf Jeff B. Pelz (pelz@cis.rit.edu)}\AND {\large \bf Christopher Kanan (kanan@rit.edu)} \AND {\large \bf Gabriel J. Diaz (gabriel.diaz@rit.edu)}\\\\
  Chester F. Carlson Center for Imaging Science\\ Rochester Institute of Technology\\ 54 Lomb Memorial Drive, Rochester, NY 14607 USA}

\begin{document}

\maketitle

\begin{abstract}

% discussion:  anticipation of the blank
% discussion:  Constraints on implementation meant that 27 ms (2 frames) was the minimum length input.

Theories for visually guided action account for online control in the presence of reliable sources of visual information, and predictive control to compensate for visuo-motor delay and temporary occlusion. In this study, we characterize the temporal relationship between information integration window and prediction distance using computational models. Subjects were immersed in a simulated environment and attempted to catch virtual balls that were transiently ``blanked'' during flight. Recurrent neural networks were trained to reproduce subject’s gaze and hand movements during blank. The models successfully predict gaze behavior within $3^{\circ}$, and hand movements within 8.5 cm as far as 500 ms in time, with integration window as short as 27 ms. Furthermore, we quantified the contribution of each input source of information to motor output through an ablation study. The model is a proof-of-concept for prediction as a discrete mapping between information integrated over time and a temporally distant motor output.

\end{abstract}

\section{Introduction}

\begin{figure}[t]
\centering
\includegraphics[width=.45\textwidth]{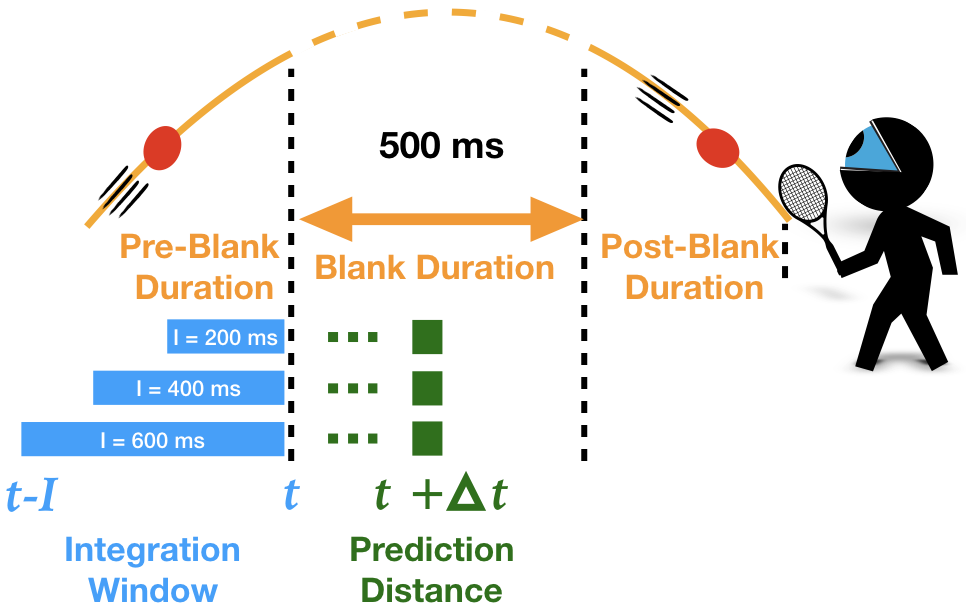}
\caption{We developed a recurrent neural network model that reproduces  human movements made in a virtual-reality ball catching task in which subjects must intercept a virtual ball that disappears for $500 \space ms$ during flight. The model integrates visual and non-visual sources of information before a blank onset from time $t-I$ through time $t$, and uses this information to reproduce the subject behavior observed at time $t+\Delta t$.  Multiple competing models were fit to the data for the purpose of exploring the minimum duration of pre-blank visual information (I) necessary to accurately reproduce behavior.}
\label{fig:modelOverview} 
\end{figure}

In the 1960's, J.J. Gibson put forth his foundational set of theories concerning \textit{ecological perception} and visually guided action~\cite{Gibson1979}.  Gibson theorized that the transformation from vision into action could be modeled as a closed-loop coupling between the parameters relevant to actions and the \textit{optical variables} that forecast a task-relevant future world state. For example, when attempting to catch a ball in flight, one can couple the time of hand closure to the ball's time to contact, which is instantaneously specified throughout the ball's approach by optical variables such as optical ${\tau}$  \cite{Lee1976,Bootsma1990,Savelsbergh1993}, equal to the ratio of the ball's instantaneous optical angle over its rate of angular expansion. Although the existence of ${\tau}$ was originally thought to underly the perception of time-to-contact across a limited range of tasks and conditions, it has since been recognized that multiple, redundant optical variables are able to provide perceptual estimates of time-to-contact, but these sources vary in their reliability across task contexts~\cite{Tresilian1999,DelaMalla2015,Lopez-Moliner2013}. The principles that determine the relative weightings placed by the perceptual system upon redundant optical variables that indicate a common task-related parameter remains a central question in the study of visually guided action. In this paper, we describe our modeling effort (as shown in Fig.~\ref{fig:modelOverview}) to elucidate these principles.
% * <jeff.pelz@gmail.com> 2018-02-01T13:16:49.020Z:
% 
% Appropriate to include the 66 reference too?
% 
% ^.

Recently, it was demonstrated that the relative weightings placed upon these redundant optical variables by the perceptual system may be partially understood through the framework of maximum likelihood estimation (MLE), which is able to account for shifts in weighting upon variables within the course of a single action~\cite{DelaMalla2015}. The authors found that reliable sources of optical information available early in the trial may influence behavior later in the trial if other reliable sources do not present themselves. A notable advantage of the MLE framework is that perceptual estimates of the task-relevant parameter (e.g., time-to-contact) may be formed through the integration of information over extended periods of time, even if they are temporally distant from the time of motor output. Thus, the model is able to capture the well-known empirical observation that, in the presence of reliable sources of information, behavior is best characterized by an online coupling with negligible latencies \cite{ZahoWarren}, as well as the finding that humans are capable of accurate short-term predictions on the basis of previously observed visual information on the order of hundreds of milliseconds \cite{Diaz2013,Zago2009}. Furthermore, there are evidence of predictive strategies aligned with ecological theory in other domains such as subjects tapping in synchrony with a metronome \cite{stephen2008strong} or walking in groups \cite{almurad2017complexity} without the need to use and internal model. However, little is known about the parameters of temporal integration, whether there exist limits to the duration over which information may be integrated, or whether there are short-term limits to the temporal distance between the integration of information and its motor output.

This study aims to further characterize the temporal characteristics of information integration using a Virtual Reality (VR) system. A head mounted display equipped with a binocular eye tracker and motion-capture systems are used to record the gaze and movement behavior of participants placed in a virtual reality catching simulator in which visual information of the ball-in-flight is unavailable for 500 ms of its parabolic trajectory (the \emph{blank period}, see Fig.~\ref{fig:modelOverview}). To investigate the temporal limits over which visual and non-visual (e.g., kinesthetic) sources of information influence the motor output, we then train multiple models, each consisting of multiple recurrent neural networks (RNNs), to reproduce the gaze positions and hand movements observed during catching. Models vary in the duration of pre-blank visual information used to predict behavior during the blank (i.e. \emph{the integration duration}), and subnetworks within a model vary in the temporal distance between the integration window and the motor output (i.e. \emph{prediction distance}). The behavior of three representative models is shown in Fig.~\ref{fig:modelOverview}. These models have integration durations of $I = \{200, 400,600\}$ and they all predict the motor output at a particular time $t+\Delta t$.

\begin{figure}[h]
   \centering
	\begin{subfigure}[a]{.4\textwidth}
        \centering
        \includegraphics[width=.9\textwidth]{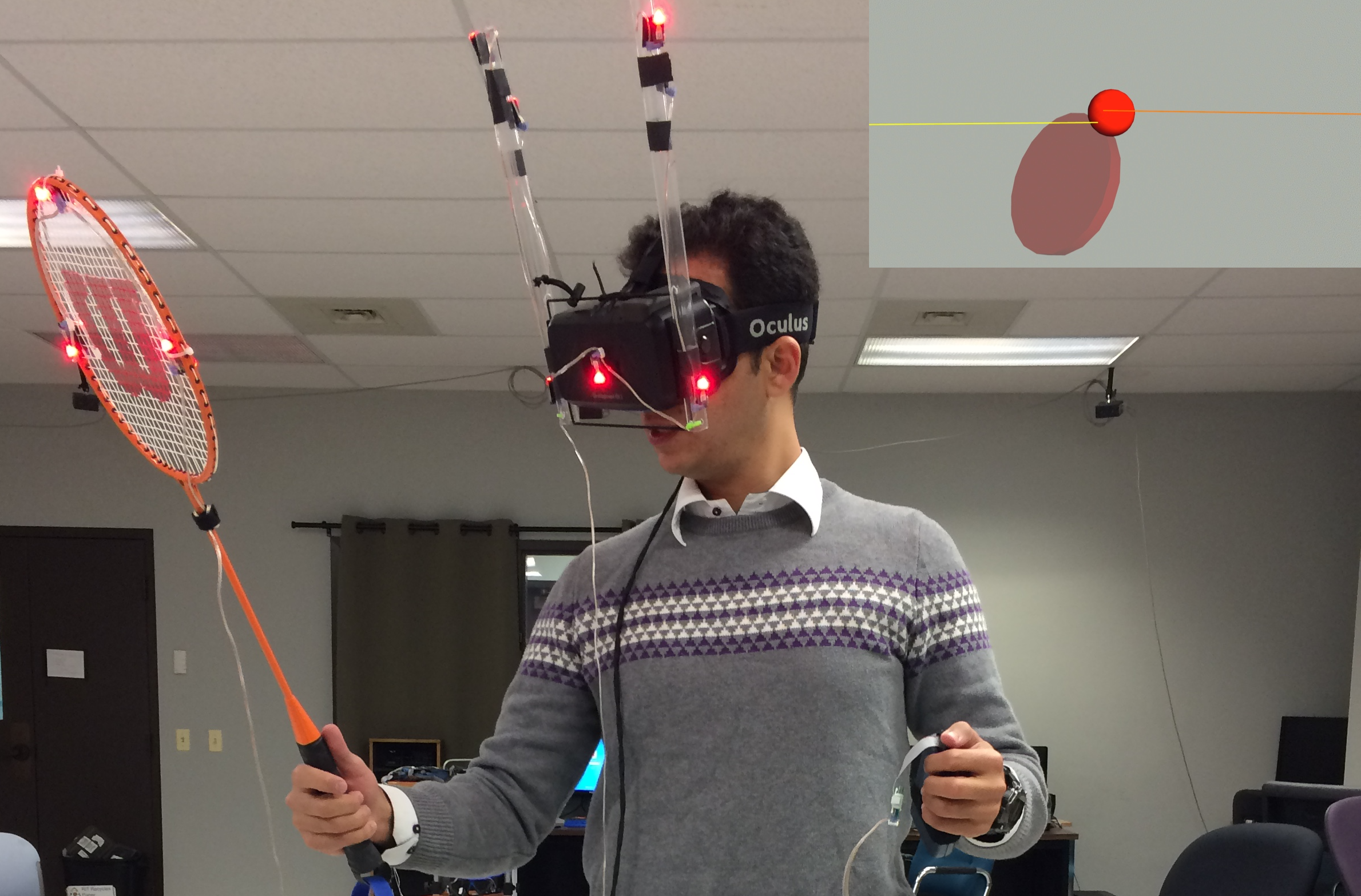}
        %\caption{}        
    \end{subfigure}
%    \begin{subfigure}[b]{.45\textwidth}
%        \centering
%        \includegraphics[width=.75\textwidth]{Figures/RearView3.png}
%    \end{subfigure}
       \caption{During data collection, subjects were immersed in a virtual ball catching task seen through an Oculus Rift DK2 with an integrated eye tracker.  Movement was tracked using motion capture markers affixed to the head mounted display (HMD) and paddle.  The scene inside the HMD is shown in the inset.  The paddle was represented as a red disc, and gaze direction by yellow/orange vectors. These vectors were not visible to the observer at the time of data collection.}
    \label{fig:theTask}
\end{figure}

\section{Procedure}
Ten participants, between 19-30 years, performed a virtual ball catching task in which they were asked to use a paddle to catch/intercept a virtual ball thrown from a distance. We use virtual reality for its ability to retain the visual structure of the natural context, while enabling us to parametrically manipulate ball trajectories.  In addition, it allowed the artificial ``blanking'' of the ball for 500 ms of its flight. This forced the subjects into a predictive mode of control during the blank for a successful catch. All participants provided informed consent prior to participation.  This study was approved by our university's Institutional Review Board, and the research was carried out in accordance with the Code of Ethics in the World Medical Association (Declaration of Helsinki). 

\subsection{Experimental Apparatus}
The VR system consists of an Intel i7-based Windows PC with two graphics cards: an NVIDIA GTX 690 driving an Oculus Rift DK2 HMD, and an NVIDIA GTX 760 driving the experimenter's desktop display. The environment was rendered using the Vizard Virtual Reality toolkit by Worldviz and physics were simulated using the OpenODE physics engine so that ball trajectories matched those expected within a real-world environment in the absence of wind resistance (Fig. \ref{fig:theTask}). Virtual imagery was presented to the subject wearing an Oculus DK2 headset with a $100^{\circ}$ field of view.  Head and paddle position/orientation were recorded at 75 Hz using a 14-camera \textit{PhaseSpace} X2 motion capture system. System latency, from the time of movement to the visual update of the screen, was measured to be less than 30 ms. Eye movements were recorded with a built-in \textit{SMI} binocular eye tracker running at 75 Hz.  A novel dynamic calibration routine was used to ensure eye tracking data was accurate throughout the experiment, and in the presence of potential helmet slippage \cite{Binaee2016}. The average eye tracking accuracy after calibration was $0.53^{\circ}$ for the central visual field ($FOV<10^{\circ}$)  and $2.51^{\circ}$ in the periphery ($10^{\circ}<FOV<30^{\circ}$). 

% At the beginning, middle, and end of the experiment, subjects participated in a calibration routine.  This routine facilitated the post-hoc calculation of a homography that minimized the angular error between the gaze vector and each calibration point.  Subsequently, this homography was subject to a temporal interpolation.

\subsection{Experimental Design}

Ball trajectories were generated using an algorithm intended to introduce sufficient variability to prevent heuristic strategies specific to the laboratory. On each trial, a red virtual ball was launched from a 6 m wide $\times$ 1.5 m high plane parallel to the virtual room's X-axis such that the ball would arrive at a 1 m $\times$ 1 m plane near the subject. Each trajectory is comprised of three durations: pre-blank, blank, and post-blank durations (see Fig.\ref{fig:modelOverview}). The pre-blank duration was randomly selected out of three values of 600, 800, or 1000 ms. The blank duration was always fixed at 500 ms, and the post-blank duration was randomly selected from 300, 400, and 500 ms values. Thus we produced 9 combinations of pre and post-blank duration, and 7 possible flight-durations. Each subject performed 135 ball catching trials. 

\section{LSTM-RNN Model of Predictive Behavior}

\begin{figure}[t]
\centering
\includegraphics[width=.45\textwidth]{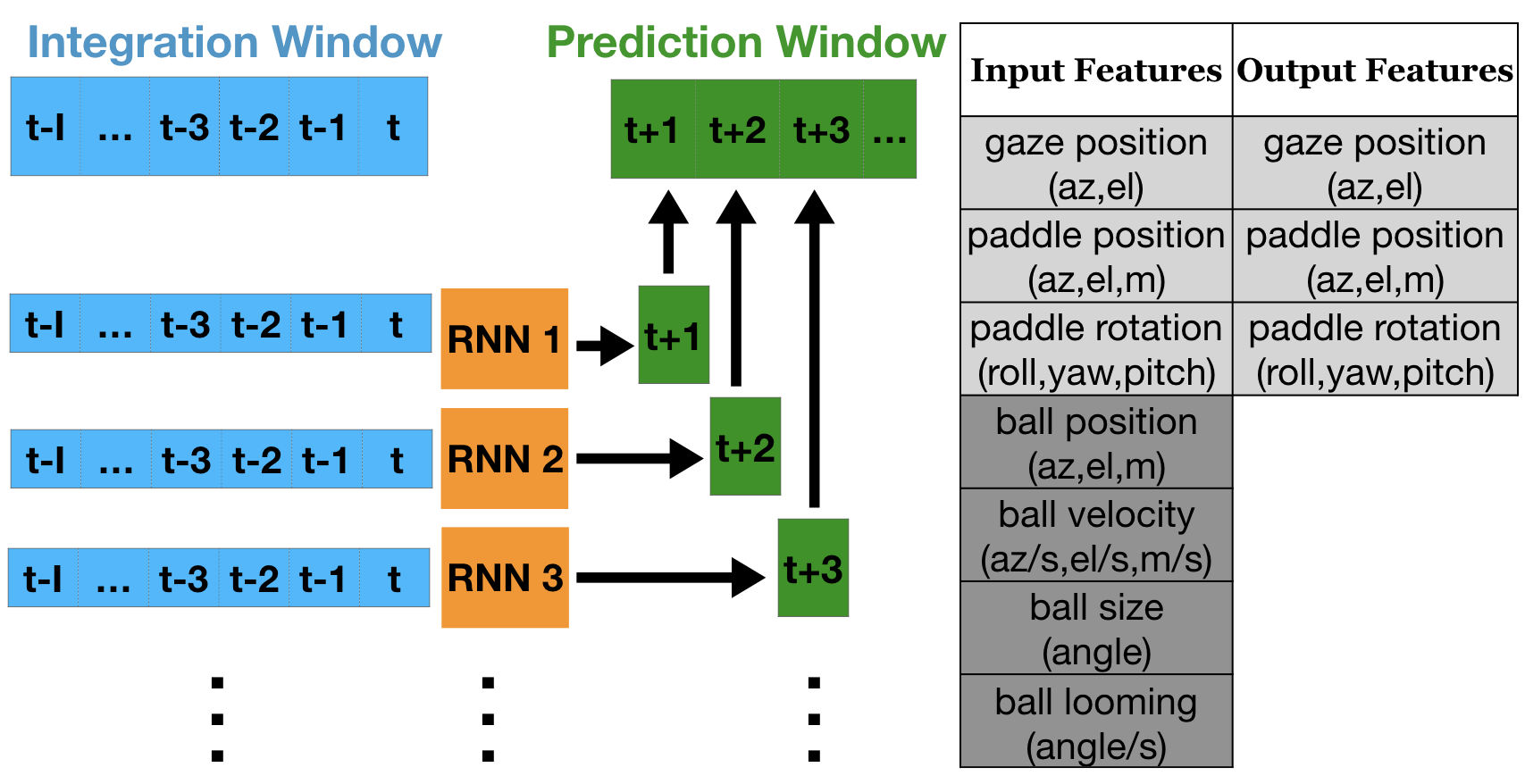}
\caption{The left panel presents a single model consisting of collection of long short-term memory recurrent neural subnetworks, each of which is responsible for predicting the motor output at a single time step during the blank. The right panel presents the input \& output feature vectors, along with their units.}
%Subnetworks in the model integrate across a window extending from $t-I$ through time $t$, where I is the integration duration, and t is the last frame before the blank. The motor output occurs at time $t+\Delta t$, where $\Delta t$ is referred to as the prediction distance. Although the integration window I is constant across subnetworks, prediction distance $\Delta t$ increases by frame increments between subnetworks up to 500 ms, the duration for which the ball is invisible during flight. On the right, we present the input \& output feature vectors, along with their units are shown here.}.
   \label{fig:theModel} 
\end{figure}

A single model of prediction across the blank duration consists of a group of long short-term memory (LSTM) subnetworks ~\cite{graves2013speech,sak2014long,sundermeyer2012lstm}. An LSTM-RNN is preferable to a simple RNN due to its robustness to exploding/vanishing gradient problems \cite{sak2014long,sundermeyer2012lstm}. A single model is presented in Fig. \ref{fig:theModel} in which each row represents an individual subnetwork. The input to each subnetwork is a sequence of visual and non-visual sources of information observed within an integration window with an integration duration $I$. The right side of the integration window is always aligned with the last frame prior to the blanking of the ball at time $t$. This means that the integration window spans from time $t-I$ through time $t$. The integration duration $I$ is constant across subnetworks belonging to a single model. The output of each subnetwork in the model is a discrete mapping from time $t$ to time $t + \Delta t$, and the prediction distance $\Delta t$ varies across subnetworks in the model. Prediction across the blank period is facilitated by monotonically increasing $\Delta t$ by frame-increments (13.3 ms) up to a duration equal to the blank period (500 ms). 

\subsection{Subnetwork Inputs and Outputs} 

The input into each subnetwork consists of a 16-dimensional input feature vectors: the first 8 dimensions corresponding to the action/motor variables, and the remaining 8 dimensions corresponding to optical variables related to the ball movement (see Fig.\ref{fig:theModel}).  Sources of information are directly calculated from the dataset geometry. The 16 element input feature vector corresponds to sources of visual information that are readily available from the stereoscopic imagery, and kinaesthetic information about the current state of the body (e.g. from proprioceptive systems, vestibular systems, and efference copy). Optical variables include ball angular position (degrees azimuth and elevation), velocity (degrees per second), ball depth from the head (meters), ball angular size (degrees), and the ball's angular rate of expansion (degrees/second). Information from kinesthesis include paddle position (meters along X, Y, and Z), paddle rotation (Euler angles roll,pitch, and yaw) and angular gaze direction (degrees azimuth and elevation). All features were defined in the head-centered, egocentric coordinate system, with the up-vector aligned with gravity, and the horizontal vector parallel with the ground plane. To normalize the feature vectors, we subtracted the mean and divided each feature by its  standard deviation, where the mean and standard deviation were computed using the entire training set. The model output is the predicted 8-dimensional action/motor state for each of the next $\Delta t$ time steps, consisting of only position and orientation information. 

%Note that the first 8 elements of input  features and output features represent the same information while the input feature elements belong to previous time steps.  \gd{This sentence was confusing, and it was ucnlear why it mattered.}

\subsection{Architecture, Training and Evaluation}

Each of the LSTM subnetworks in the model has 1 hidden layer of 25 LSTM cells. In preliminary experiments, we did not observe improvements using additional cells. The hidden layer of each LSTM projects to a fully connected layer with 8 units that predict a future motor/action state. Because training was meant to account for predictive behavior, and not online control, we restricted training to periods in which the motor output occurred during the blank. Each model has 37 rows of subnetworks, hence each subnetwork is responsible for predicting the motor/action state at time $\Delta t$, where values range from 13.33 ms to 493.33 ms into the future, with a resolution of 13.33 ms.

We split the dataset into train (68\%), validation (12\%), and test (20\%) partitions. The model was trained on all 135 trials of all 10 subjects. The models were optimized using the Adam optimizer with a learning rate of 0.0001 and the settings recommended in \cite{kingma2014adam}, e.g., batch size of 128 and 2000 epochs. Early stopping based on validation loss was also used with patience parameter set to 100. The dataset is formatted into Pandas data frame and is available as an online repository.

\section{Results}

Subjects on average caught the ball on 67\% of trials (SD: 14\%). During the blank period, the invisible ball moved between 10.3 degrees (pre-blank: 600, post-blank: 500 ms), and 12.6 degrees (pre-blank: 1000ms, post-blank: 300 ms) through the subject's visual field. During the blank, subjects tracked the ball through coordinated movements of the eyes and head.  The ratio of angular displacement of gaze over that of the ball reveals that subjects accounted for 0.95 of the ball's displacement across all conditions (SD=0.11; t(9)=-1.43, \emph{p}=0.187). Upon reappearance, the ball was moving approximately 34.1 degrees per second (SD: 4.3), and the gaze vector was well matched to the ball's angular velocity, as indicated by a pursuit gain (ratio of angular velocity of the ball over gaze) of 0.94 (SD: 0.11; t(9)=-1.14, \emph{p}=0.28). There were also variations with the timing of the blank e.g., with variation of the pre-blank duration and post-blank duration.

\subsection{Model performance}

\begin{figure}[h]
\centering
\includegraphics[width=.5\textwidth]{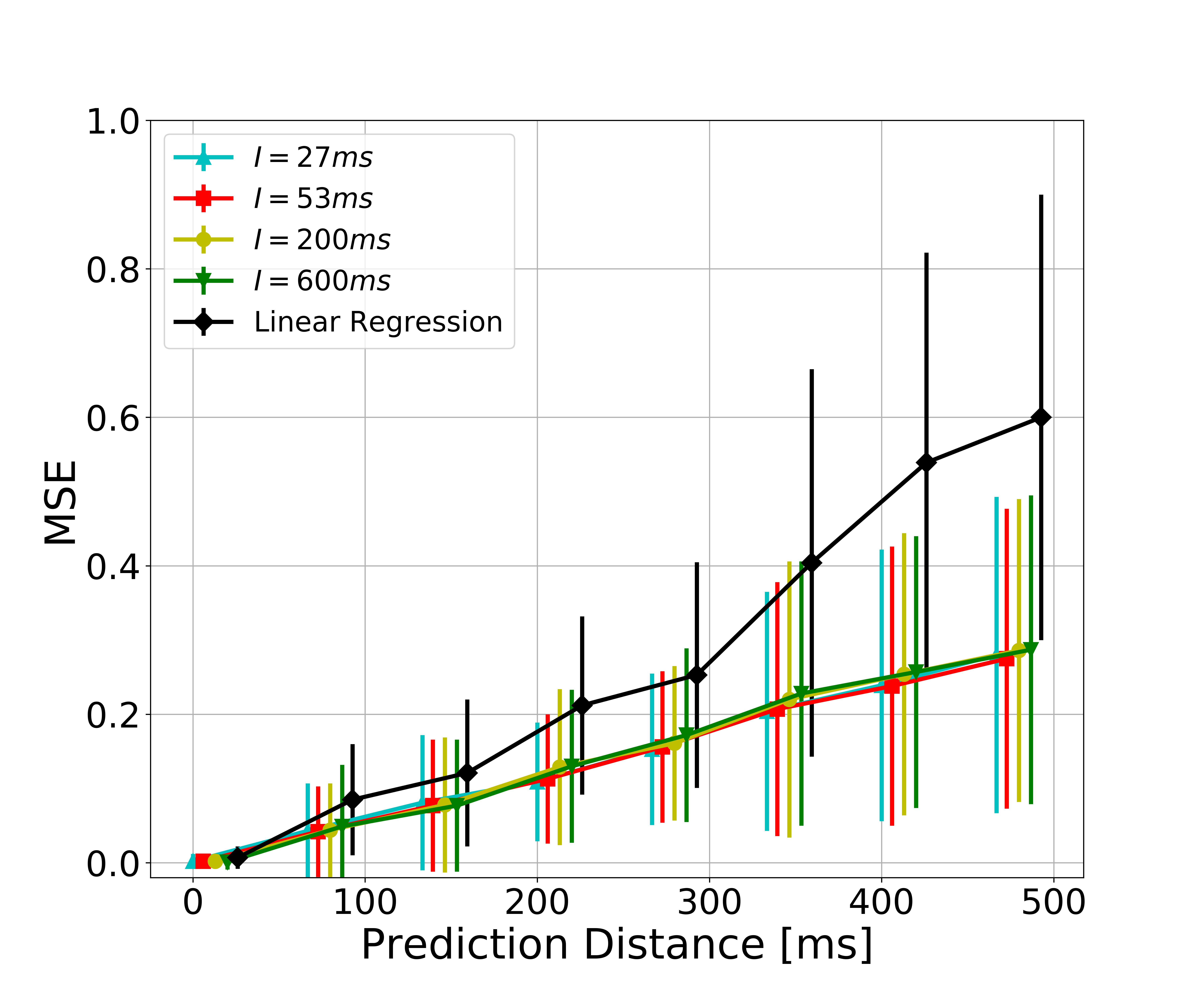}
\caption{The performance of four models differing by integration duration ($I=\{27$, $53$, $200$, and $600\}$ ms) as predictors of motor output throughout the $500 \space ms$ blank period. For comparison, we also include a linear regression based upon the sensory evidence available before the blank.}
   \label{fig:errorByInegrationDur} 
\end{figure}

Fig.\ref{fig:errorByInegrationDur} presents the mean-squared error (MSE) for four models ($I=\{27$, $53$, $200$, and $600\}$ ms) when predicting gaze and motor behavior throughout the blank period.  For reference, we also indicate the results of the linear regression between information prior to the blank period and the motor output. For all models, both MSE and variability increased with prediction distance. The four LSTM models outperformed the linear regression by a magnitude that grows with prediction distance. The observation that there is no added benefit to increasing the integration duration beyond 27 ms suggests that 27 ms of visual information prior the blank is sufficient to account for the predictive movements observed during the blank.

\begin{figure}[t]
\centering
   \begin{subfigure}[b]{0.55\textwidth}
	   \includegraphics[width=.96\textwidth]{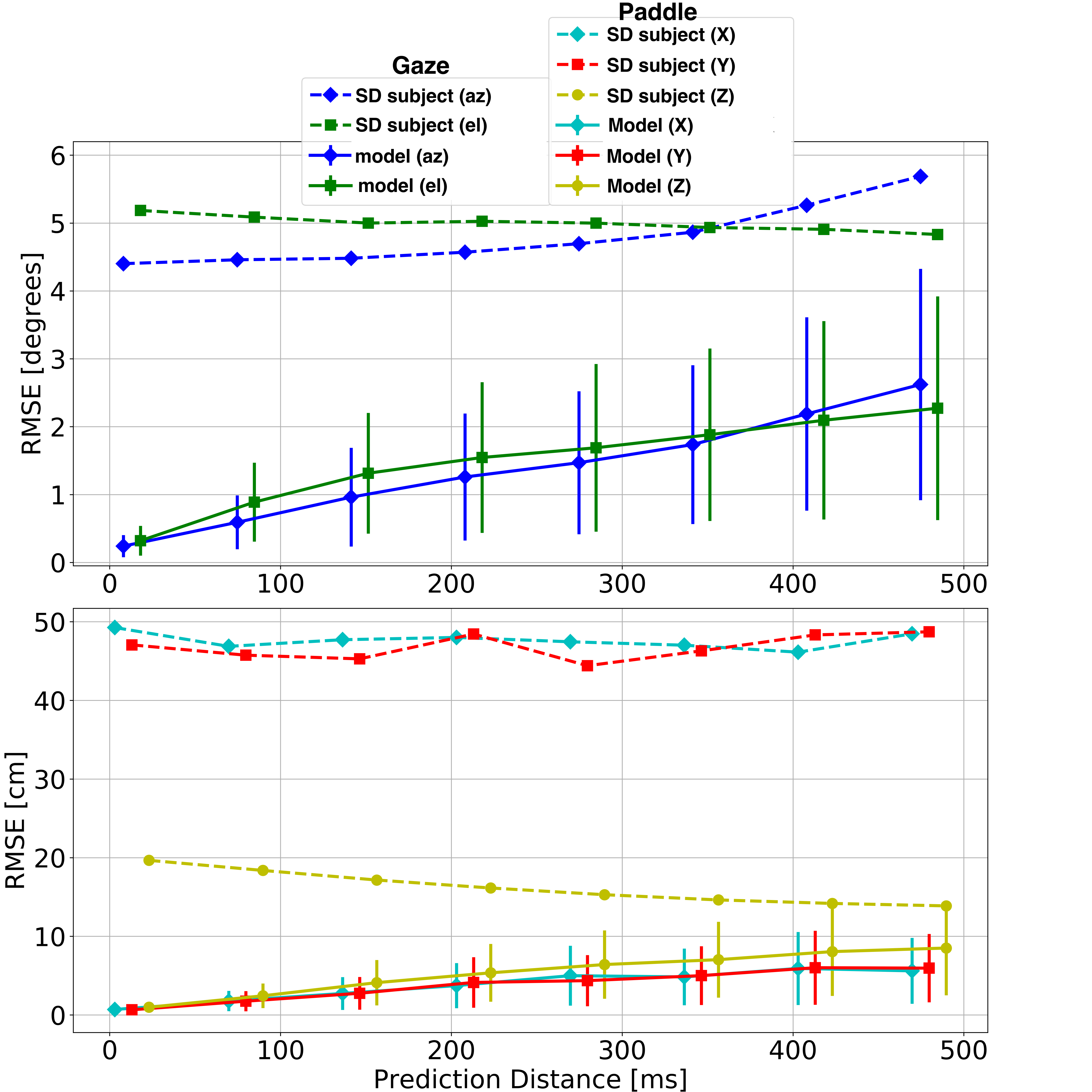}
       %\caption{}
       \label{fig:gaze_prediction} 
	\end{subfigure}    
    \caption{Root-mean squared error (RMSE) for the azimuth and elevation of the gaze vector in head-centered polar coordinates (top panel), and paddle position relative to the head in metric coordinates (bottom panel) for the model with $I=\space 27ms$. Dotted lines represent the standard deviation of the subject's gaze vector from the mean.  These values provide an estimate of the RMSE expected for a model with an output equal to the per-frame mean gaze direction, and that does not account for trial-by-trial variations in the ball's trajectory.}
\label{fig:gazeHandPrediction}
\end{figure}

Fig.~\ref{fig:gazeHandPrediction} shows the root mean squared error for the model with $I$=27ms as a predictor of gaze position (Fig.~\ref{fig:gazeHandPrediction} top) and paddle position (Fig.~\ref{fig:gazeHandPrediction} bottom). In addition, dotted lines represent the the standard deviation of subject data around the grand mean, which is an estimate of the amount of unaccounted variance one would expect from a model that simply estimates the mean value at each frame of the blank period. The observation that model RMSE is lower than this estimate is evidence that the model is able to account for trial-by-trial variations in ball trajectory on the basis of visual and kinesthetic input features. 

\subsection{Visual prediction, or a simple motor-to-motor mapping?}

% \subsection{Does the model rely upon ball visual information, or a simple motor-to-motor mapping?}

\begin{figure}[htb]
\centering
\includegraphics[width=.5\textwidth]{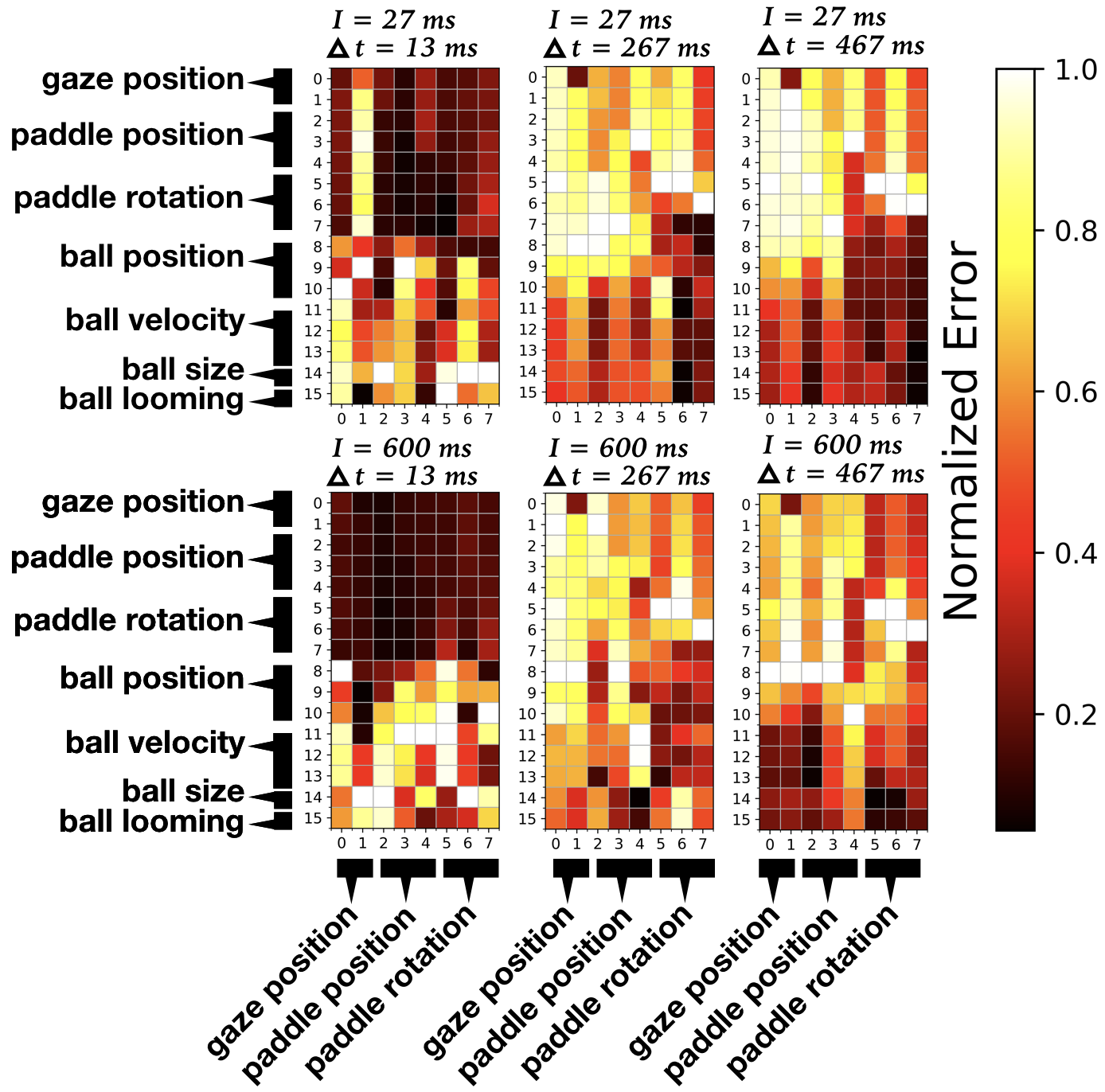}
\caption{To test the relative contribution of input features to the accurate reproduction of the observed motor output, features were iteratively removed following training.  Here, we present the resulting mean error in covariance following iterative input feature ablation for two values of $I = \{27,600\} ms$ and three values of $\Delta t = \{13, 267, 467\} ms$. Rows indicate which feature was removed, columns correspond to the output feature, and brightness indicates the magnitude of error in covariance as a result of feature ablation.}
   \label{fig:FeatureSignificance} 
\end{figure}

Although LSTM-based models of visual prediction outperformed linear regression as a predictor of gaze and motor behavior throughout the blank period, measurements of error alone cannot rule out the possibility that the model was performing a simple extrapolation of motor variables, while disregarding the visual input state of the environment. To investigate, we ran a series of iterative tests in which individual input features were systematically removed from the subnetworks, and monitored the performance of subnetworks responsible for output at different stages of the blank period. The models/subnetwork training regime was not altered between tests, and included the full range of inputs. The assumption is that the ablation of an important input feature following training would result in an increase in mean reproduction error proportional to the feature's influence on the model's ability to reproduce the observed motor outputs. The results of these iterative ablation studies are presented in Fig.~\ref{fig:FeatureSignificance} for two models  $I = \{27,600\} ms$, and for three prediction distances  ($\Delta t = \{13, 267, 467\} ms$). To account for differences in units, the error values indicated by cell brightness were max-normalized across the output features represented by columns.

By comparing between rows within a single panel in Fig.~\ref{fig:FeatureSignificance}, one can visually compare the relative contribution of visual features (e.g. ball position, velocity, angular size, and looming) and features related to the subnetwork's motor output (e.g. gaze position, paddle position, paddle rotation) to movement reproduction. For example, in the bottom-left panel ($I = 600$ ms, $\Delta t =$ 13ms), it is clear that the subnetwork relied heavily upon visual sources of information concerning the state of the ball for the accurate reproduction of the observed motor outputs. Removing ball visual features caused on average 31\% more error compared to gaze\&paddle position/rotation.  This suggests that, when integration time is long, visual information concerning the ball's trajectory is the best indicator of the motor behavior observed over short distances. There is a similar result when $I$=27, with the exception that the ablation of motor variables (the upper half of rows in Fig.~\ref{fig:FeatureSignificance}) degraded the reproduction of gaze elevation (column \#1). The results of this ablation study suggest that, despite the lack of a benefit of increased integration time to overall RMSE, this may result in an increased robustness following the loss of an expected input feature.

Comparison across different prediction distances (between left, middle and right figures in Fig. \ref{fig:errorByInegrationDur}) suggests that, when predicting further in time, one must rely upon a combination of input features related to the visual and motor state. This is true regardless of integration time, although values suggest a slight bias (less than 8\%) towards motor variables when integration time is low (in the top middle and top right panels of Fig. \ref{fig:FeatureSignificance}).

\section{Discussion and Conclusions}

In this study, we trained a series of competing models to reproduce the gaze and motor behavior made by subjects performing a catching task in which the virtual ball was transiently blanked for a portion of its flight.  Under the constraints imposed by the task, only 27 ms of visual and kinesthetic information prior to the occlusion was necessary to accurately reproduce up to 500 ms of behavior following the removal of sensory feedback (Fig \ref{fig:errorByInegrationDur}). Despite the low integration time of 27 ms, our models were able to predict gaze position within 3\degree of accuracy almost 500 ms after the ball's disappearance. This value is far below that expected by a model that simply estimates the time-varying mean, suggesting that the model was able to account for trial-by-trial behavior variations in response to changes in ball trajectory. Similarly, the model was able to reproduce hand position within 8.5 cm of error, 500 ms after the ball's disappearance (Fig.\ref{fig:gazeHandPrediction}). Although overall model performance did not vary with changes in integration duration, we found that the models ability to reproduce temporally distant behavior (i.e. at higher values of $\Delta t$) required input from both visual information and kinesthetic sources of information. The results of this task provide further insight into the temporal dynamics between sensory information and the motor output over the course of a single action.

The low-error observed for our model at prediction distances near to 500 ms provides evidence against the argument that accurate prediction requires internal models of physical dynamics (e.g. of Newton's law) for the continuous extrapolation of the ball's trajectory following occlusion \cite{Zago2009}. Instead, our models learned temporally discrete mappings between evidence integrated from time $t-I$ through $I$, and a motor output at temporally discrete time in the future (time $t+\Delta t$). The prediction distance at 500 ms in duration was roughly half what would be expected by a model that simply predicts the mean motor state (Fig. \ref{fig:gazeHandPrediction}), suggesting that such a simple mapping could be sufficient if one presupposes the availability of the information sources included in the model inputs. Moreover, it is notable that all variables were specified within a head-centered, ego-centric reference frame, and did not presuppose reconstruction of the visual surroundings within a Euclidean frame of reference. Finally, by systematically exploring the error introduced by the ablation of input features in serial, we provided evidence against the possibility that the model was simply learning temporal correlations between subsequent motor states in the absence of visual input concerning ball position.  Thus, the model serves as a proof-of-concept for the possibility that visual-motor prediction is a temporally discrete mapping between previously observed world states and a temporally distant motor output.

% Limitations of the model
It is notable that the biological organism is subject to additional constraints not considered in the current architecture.  Most notably, the model in no-way accounts for the influence of perceptual processing, or the perceptual sensitivities of the human visual system that further influence the reliability of information sources over time ~\cite{cutting1995potency}. Similarly, the model does not account for short-term decay in visual working memory \cite{Issen2012}. Finally, due to limitations in the LSTM-RNN framework preventing the use of dynamically sized integration durations within a single model, our model was unable to account for its own output motor states between the time of ball blanking (time $t$) and the current motor output (time $t+\Delta t$). Although our approach does demonstrate that 27 ms of sensory information is sufficient to explain predictive subject behavior, the human visual system must undoubtedly integrate across longer durations to overcome these biological constraints. The influence of these constraints might be explored in future work, for example, through systematic degradation of the visual input to reflect the constraints imposed by early visual processing.

\section{Acknowledgments}
The authors would like to thank Dr. David W. Messinger for his support and Rakshit S. Kothari for consultation, help and discussions during this study.

\bibliographystyle{apacite}

\setlength{\bibleftmargin}{.125in}
\setlength{\bibindent}{-\bibleftmargin}

\bibliography{cogSci.bib}

\end{document}